\documentclass[journal]{IEEEtran}
\usepackage[numbers,sort&compress]{natbib}
\usepackage{amsmath,amssymb,amsfonts}
\usepackage{algpseudocode}
\usepackage{algorithmicx,algorithm}
\usepackage{graphicx}
\usepackage[caption=false,font=footnotesize,labelfont=rm,textfont=rm]{subfig}
\usepackage{xcolor}
\usepackage{bm}
\usepackage{stfloats}
\usepackage{makecell}
\usepackage{booktabs}
\usepackage{multirow}
\usepackage{cleveref}
\usepackage[table]{xcolor}
\usepackage{threeparttable}
\usepackage{empheq}
\usepackage[theorems,skins]{tcolorbox}
\usepackage{ragged2e}
\begin{document}
\newtheorem{proposition}{Proposition}
\newtheorem{lemma}{Lemma}
\newtheorem{corollary}{Corollary}

\title{WiFo-INR: A Wireless Foundation Model Based on Implicit Neural Representations}

\author{Boxun Liu,~\IEEEmembership{Graduate Student Member,~IEEE,}
Xuanyu Liu,~\IEEEmembership{Graduate Student Member,~IEEE,}\\
Shijian Gao,~\IEEEmembership{Member, IEEE,} Xiang Cheng,~\IEEEmembership{Fellow,~IEEE,}
Liuqing Yang,~\IEEEmembership{Fellow,~IEEE}

\thanks{
Boxun Liu, Xuanyu Liu, and Xiang Cheng are with the State Key Laboratory of Photonics and Communications, School of Electronics, Peking University, Beijing 100871, China (e-mail: boxunliu@stu.pku.edu.cn; xyliu25@stu.pku.edu.cn; xiangcheng@pku.edu.cn).

Shijian Gao is with the Internet of Things Thrust, The Hong Kong University of Science and Technology (Guangzhou), Guangzhou 511400, China (e-mail: shijiangao@hkust-gz.edu.cn).

Liuqing Yang is with the Internet of Things Thrust and Intelligent Transportation Thrust, The Hong Kong University of Science and Technology (Guangzhou), Guangzhou 511400, China (e-mail: lqyang@ust.hk).

}}

\maketitle
\begin{abstract}
Wireless foundation models are emerging as a promising paradigm for AI-native physical-layer design. 
However, existing methods typically model channel state information (CSI) as image-like discrete tensors with generic token decoders that may struggle to capture complex high-frequency variations efficiently and often produce high-dimensional, size-dependent representations.
In this paper, we propose WiFo-INR, an implicit neural representation (INR)-based wireless foundation model that represents CSI as a coordinate-conditioned neural function. 
A Transformer encoder maps partial or coarse CSI to fixed-dimensional modulation tokens that adapt a SIREN-based decoder, and a compression autoencoder enables quantized CSI feedback. 
It adopts a two-stage self-supervised pretraining scheme, where mixed masking and denoising improve channel reconstruction and compression-enhanced pretraining enables accurate CSI feedback at low compression ratios. 
Extensive experiments demonstrate that WiFo-INR learns efficient, compact, and CSI-size-independent implicit wireless representations. 
Compared with existing foundation models, WiFo-INR improves channel reconstruction and CSI feedback performance while substantially reducing inference latency.
It also transfers efficiently to diverse wireless tasks with minimal fine-tuning overhead and achieves zero-shot generalization to unseen CSI sizes.
\end{abstract}


\section{Introduction}
\IEEEPARstart{T}{he} forthcoming sixth-generation (6G) wireless networks are expected to support ubiquitous intelligent agents and emerging applications such as collaborative embodied intelligence. 
This vision requires broader coverage, including satellite communications, diverse configurations such as millimeter-wave bands and massive antenna arrays, and multimodal sensing capabilities for Synesthesia of Machines (SoM) \cite{cheng2024intelligent}. 
The resulting heterogeneity poses substantial challenges to physical-layer design. Existing algorithms largely rely on model-driven parametric methods whose performance depends on accurate prior knowledge. 
In realistic propagation environments, however, difficult-to-model non-idealities often cause model mismatch and limit their generality.

Over the past decade, AI has been widely adopted in physical-layer design to improve performance and robustness through data-driven modeling. 
However, most existing approaches remain task-specific, with separate models required for different tasks, system configurations, and scenarios. 
As a result, the number of models deployed on a communication device grows with their combination, leading to substantial overhead in storage, training, management, and data collection. This paradigm therefore scales poorly in heterogeneous wireless systems.
\begin{figure}[!t]
    \centering
    \includegraphics[width=0.9\linewidth]{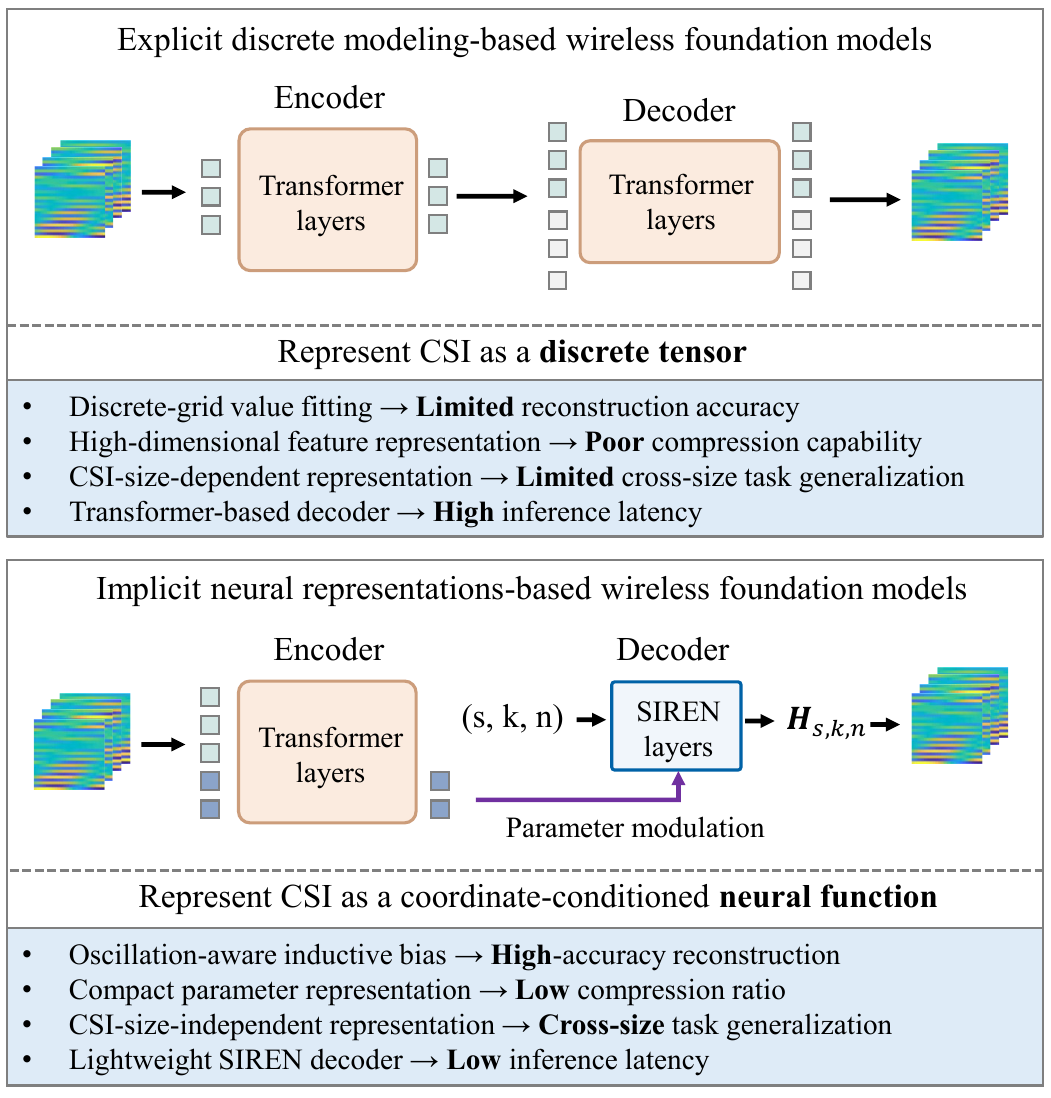}
    \vspace{-2mm}
    \caption{Comparison of wireless foundation model architectures based on explicit discrete modeling and INRs.}
    \vspace{-6mm}
    \label{paradigm comparison}
\end{figure}
In recent years, foundation models \cite{bommasani2021opportunities} have driven a paradigm shift in AI. Inspired by this success, domain-specific foundation models for wireless systems, termed wireless foundation models \cite{yang2025wirelessgpt,catak2025bert4mimo,zheng2025muse,alikhani2024large,liu2025wifo,guler2026multi,jiang2025mimo,pan2025large,
zhang2026adaptive,bian2026airfm,liu2026wifo,liu2025WiFo2,liu2026wifomisac}, have been proposed to address the aforementioned challenges. 
By pretraining on large-scale wireless data, typically channel state information (CSI), these models learn general-purpose wireless representations \cite{cheng2026large,cheng2025som,liang2026large,guo2026large} that transfer across scenarios, system configurations, and tasks, enabling few-shot or even zero-shot adaptation to a broad range of wireless downstream tasks. 
For example, the early study LWM \cite{alikhani2024large} obtained general representations through self-supervised masked channel modeling on spatial-frequency CSI data and can efficiently adapt to multiple downstream tasks such as beam prediction. 
The contemporaneous work WiFo \cite{liu2025wifo} performed hybrid masked reconstruction pretraining on heterogeneous spatial-temporal-frequency CSI and, for the first time, achieved zero-shot channel prediction that outperformed fully supervised training of task-specific models. 
Subsequent studies further improved pretraining objectives and network architectures, for example by introducing contrastive learning \cite{guler2026multi,jiang2025mimo,pan2025large}, three-dimensional rotary position embedding (3D-RoPE) \cite{zhang2026adaptive}, and modeling in the delay-Doppler-angle (DDA) domain \cite{bian2026airfm}, thereby enhancing the modeling and generalization capabilities of general-purpose wireless representations. 
In addition, some studies have incorporated specialized designs that bias networks towards particular classes of wireless tasks, such as CSI feedback \cite{liu2026wifo}, achieving strong zero-shot generalization. 
More recently, WiFo-2 \cite{liu2025WiFo2} has enabled the unified processing of 12 wireless communication and sensing tasks, while substantially reducing the fine-tuning overhead.

However, existing wireless foundation models largely inherit a discrete-tensor modeling paradigm from image processing and represent CSI on fixed space-time-frequency grids. 
Such tensors are finite samples obtained under specific bandwidth, subcarrier spacing, antenna configuration, and temporal sampling settings. 
This is a valid abstraction, but CSI samples exhibit structured oscillatory variations induced by multipath propagation. 
Generic token-based decoders are not explicitly tailored to this frequency structure and may struggle to capture complex high-frequency channel variations efficiently. 
In addition, their high-dimensional features can increase feedback overhead, while the number of feature tokens changes with CSI size and restricts direct reuse across input scales. 
Transformer decoders also reconstruct dense CSI grids through repeated token interactions, which incurs considerable latency. 
These observations motivate a coordinate-conditioned alternative with a frequency-aware sinusoidal inductive bias and compact, fixed-dimensional instance representations.

Implicit neural representations (INRs) \cite{sitzmann2020implicit} offer a new opportunity to overcome these bottlenecks. 
An INR represents a signal as a coordinate-conditioned neural function that maps coordinates to signal values, making it well suited to capturing complex wireless channel variations while yielding compact parameter-based representations.
Several studies have begun applying INRs to wireless tasks such as RF propagation-field reconstruction \cite{zhao2023nerf2}, CSI feedback \cite{wu2024mimo}, and channel estimation \cite{shi2026unsupervised}, providing early evidence of their potential for efficient wireless channel representation. 
For example, \cite{wu2024mimo} first introduced INRs for massive multiple-input multiple-output (MIMO) CSI feedback by representing CSI as a neural function of antenna and subcarrier coordinates. It optimizes, at the user side, an instance-specific modulation codeword for each CSI instance to enable feedback, thereby achieving CSI reconstruction at extremely high compression ratios. 
In \cite{shi2026unsupervised}, SIREN was used to represent high-mobility OFDM channels as INRs, where the INR parameters are optimized from the received signals in the current slot to achieve robust channel estimation. 
Nevertheless, these methods still require gradient-based optimization for each CSI instance, resulting in substantial online training latency.
Moreover, existing efforts are often tailored to a single wireless task, making it difficult to develop a unified design for diverse wireless tasks.

To address these limitations, we propose WiFo-INR, an INR-based wireless foundation model that learns generalizable implicit wireless representations. 
As shown in Fig. 1, WiFo-INR exploits the inductive bias that grid-sampled CSI exhibits structured oscillatory variations induced by multipath propagation and models each instance with a coordinate-conditioned neural function. 
Specifically, we develop an architecture that generalizes across CSI instances and comprises three key modules. 
A Transformer-based encoder maps each CSI instance to modulation tokens that adapt the decoder, a SIREN-based decoder provides frequency-aware coordinate mapping over the three CSI dimensions, and a compression autoencoder further compresses the tokens. 
We also design a two-stage hybrid self-supervised pretraining scheme. 
The first stage combines mixed masking and interpolation-denoising tasks to enhance 3D CSI estimation and prediction, while the second stage employs compression-reconstruction pretraining to enable 2D CSI feedback at extremely low compression ratios. 
The main contributions are summarized as follows:
\begin{itemize}
\item We propose WiFo-INR, to the best of our knowledge the first INR-based wireless foundation model, which uses coordinate-conditioned neural functions to model sampled CSI. 
It provides a unified solution for channel acquisition, including channel estimation, channel prediction, and CSI feedback, while learning compact and general-purpose implicit representations for diverse wireless communication and sensing tasks.

\item We develop a cross-instance generalizable architecture in which a SIREN-based decoder represents CSI as an INR and a Transformer-based encoder generates modulation tokens to adapt the decoder to each CSI instance. 
We further design a two-stage self-supervised pretraining scheme to improve channel reconstruction accuracy and support CSI feedback at low compression ratios.

\item Extensive experiments demonstrate WiFo-INR's superiority across diverse wireless tasks. 
Compared with existing wireless foundation models based on explicit discrete modeling, WiFo-INR improves channel reconstruction and CSI feedback performance while substantially reducing inference costs. 
It also transfers efficiently to CSI-related downstream tasks, including wireless localization, beam prediction, and scenario classification, with minimal fine-tuning overhead, and achieves zero-shot generalization across different CSI sizes.
\end{itemize}

\textit{Notations}: $\bm{a}_i$ is the $i$-th element of a vector $\bm{a}$, $\bm{A}_{i,j}$ denotes the element of matrix $\bm{A}$ at the $i$-th row and the $j$-th column, and $\bm{A}_{:,m}$ denotes the $m$-th column of $\bm{A}$.
$\lVert \cdot \rVert_{F}$ denotes the Frobenius norm.
$\mathbb{R}$ and $\mathbb{C}$ denote the sets of real numbers and
complex numbers, respectively.

\section{System Model and Problem Formulation}
In this section, we first introduce the system model and formulate CSI-related wireless tasks. 
We then present wireless foundation models based on explicit discrete modeling and discuss their limitations.
\subsection{System Setup}
We consider a MIMO-orthogonal frequency division multiplexing (OFDM) system where the base station (BS) is equipped with a uniform planar array (UPA), and the user is equipped with a single antenna. 
Let $N_h$ and $N_v$ denote the numbers of antenna elements along the horizontal and vertical dimensions of the UPA, respectively.
We adopt a classical geometric channel model with $P$ propagation paths. 
For the $p$-th path, the complex path gain, Doppler frequency shift, delay, azimuth angle, and elevation angle are denoted by $\beta_p$, $\mu_p$, $\tau_p$, $\theta_p$, and $\phi_p$, respectively. 
The spatial CSI function $\bm{h}(t,f)\in\mathbb{C}^{N_hN_v}$ between the BS and the user at time $t$ and frequency $f$ is then given by
\begin{align}\label{h-tf}
\bm{h}(t,f)=\sum_{p=1}^{P}\beta_p e^{j2\pi \mu_p t} e^{-j2\pi f \tau_p}\bm{a}(\theta_p,\phi_p), 
\end{align}
where $\bm{a}(\theta,\phi)\in\mathbb{C}^{N_hN_v}$ represents the corresponding array steering vector of the $p$-th path.

Given $T$ sampling points in the time domain with sampling interval $\Delta t$, and $K$ subcarriers in the frequency domain with subcarrier spacing $\Delta f$, the time–frequency-space CSI tensor observed over the resulting three-dimensional discrete grid is denoted as $\bm{H}\in\mathbb{C}^{T \times K \times N}$, where $N=N_hN_v$. 
For any three-dimensional coordinate $(s, k, n)$ within the domain, the following relation holds: 
\begin{align}\label{H-tkn}
    \bm{H}_{s,k,n}= [\bm{h}\left(s\Delta t,f_0+k\Delta f\right)]_n,
\end{align}
where $f_0$ represents the frequency of the first subcarrier.

\subsection{Overview of CSI-Related Tasks}
CSI plays a central role in wireless communication and sensing tasks. Here, we consider the following five categories of CSI-related tasks. 
In this paper, we define channel reconstruction as the process of recovering complete CSI from partial CSI, encompassing channel estimation, time-domain channel prediction, and frequency-domain channel prediction. 
In addition, we use channel acquisition as an umbrella term for channel reconstruction and CSI feedback tasks. 
\subsubsection{Channel Estimation}
Channel estimation aims to reconstruct the complete 3D CSI tensor from the observed signals at a subset of pilot positions. 
Since neural networks have difficulty efficiently handling diverse pilot patterns, existing AI-based designs typically first employ a least-squares (LS) estimator to estimate the CSI at the pilot positions and then use linear interpolation to obtain a coarse CSI estimate, denoted by $\bm{\bar H}$. 
The AI model then further processes $\bm{\bar H}$ to produce a refined estimate, denoted by $\bm{\hat H}$. 
Denote the corresponding mapping function as $f_{\rm CE}$ and the process is given as
\begin{align}
\bm{\hat H} = f_{\rm CE} (\bm{\bar H}).
\end{align}

\subsubsection{Time-Domain Channel Prediction}
Time-domain channel prediction aims to predict the CSI at the next $T-T_{\rm h}$ time instants from the CSI observed at the preceding $T_{\rm h}$ time instants, thereby mitigating channel aging caused by high mobility of the transmitter and receiver. 
Let $f_{\rm TCP}$ denote the mapping function, and this process can be modeled as
\begin{align}
\bm{\hat H}_{T_{\rm h}+1:T,:,:} = f_{\rm TCP} (\bm{H}_{1:T_{\rm h},:,:}).
\end{align}

\subsubsection{Frequency-Domain Channel Prediction}
Frequency-domain channel prediction aims to predict the CSI at the subsequent $K-K_{\rm u}$ neighboring subcarriers from the CSI observed at the preceding $K_{\rm u}$ subcarriers, thereby substantially reducing pilot overhead. 
Let $f_{\rm FCP}$ denote the mapping function, and this process is defined as
\begin{align}
\bm{\hat H}_{:,K_{\rm u}+1:K,:} = f_{\rm FCP} (\bm{H}_{:,1:K_{\rm u},:}).
\end{align}

\subsubsection{CSI Feedback}
CSI feedback aims to compress space-frequency two-dimensional CSI into low-dimensional information and transmit it to the receiver side for reconstruction. 
A typical scenario arises in frequency division duplex (FDD) systems or time division duplex (TDD) systems with imperfect channel reciprocity, where the user measures the downlink CSI and feeds it back to the BS for precoding. Let $f_{\rm C}$ and $f_{\rm R}$ denote the compressing and recovery mapping functions, and this process is defined as
\begin{align}
\bm{\hat H}=f_{\rm R}(f_{\rm C}(\bm{H})).
\end{align}
Note that, in this case, the time dimension of $H$ has size 1.

\subsubsection{CSI-Related Downstream Tasks}
In wireless systems, there exist multiple other CSI-based communication and sensing tasks, which can be uniformly modeled as classification or regression problems with CSI as the input. 
Let $f_{\rm down}$ denote the mapping function and $\bm{r}_{\rm d}$ denote the output, and this process is given by
\begin{align}
\bm{r}_{\rm d} = f_{\rm down} (\bm{H}).
\end{align}

\subsection{Wireless Foundation Models based on Explicit Discrete Modeling}
The aforementioned CSI-related tasks can be broadly divided into two categories: one that estimates complete CSI from partial and noisy CSI observations or compressed features, and another that performs CSI-based classification or regression. 
This formulation is naturally aligned with the architecture and pretraining paradigm of masked autoencoders (MAE), thereby enabling the development of a wireless foundation model capable of handling diverse CSI-related tasks. 
Consider a typical MAE-based wireless foundation model, such as WiFo, which consists of a Transformer-based encoder and decoder. 
The encoder maps the estimated partial CSI $\bm{\tilde H}_{S}$ into a sequence of feature tokens $\bm{T}_{\rm F}\in\mathbb{R}^{N_{\rm F} \times D_{\rm F}}$, which are then combined with learnable mask tokens $\bm{T}_{\rm M}\in\mathbb{R}^{N_{\rm M} \times D_{\rm F}}$ and fed into the decoder to reconstruct the complete CSI $\bm{\hat H}$. 
Here, $S$ denotes a subset of the full element set $\Omega$, $D_{\rm F}$ denotes the dimensionality of tokens, and $N_{\rm F}$ and $N_{\rm M}$ denote the number of feature tokens and mask tokens, respectively.
The overall process can be expressed as
\begin{align}
\bm{\hat H} = f_{\rm dec} (\bm{T}_{\rm F}, \bm{T}_{\rm M}), \ \bm{T}_{\rm F}= f_{enc} (\bm{\tilde H}_{S}),
\end{align}
where $f_{enc}$ and $f_{\rm dec}$ denote the encoding and decoding processes, respectively.
Specifically, when $S$ corresponds to the full set, the time-domain subset, and the frequency-domain subset, it naturally instantiates the tasks of channel estimation, time-domain channel prediction, and frequency-domain channel prediction, respectively.
Moreover, at a high masking ratio, $\bm{T}_{\rm F}$ itself can serve as a compact feature representation that can be quantized and used for compressed feedback. 
In addition, $\bm{T}_{\rm F}$ can be regarded as a general-purpose representation and further fine-tuned for downstream tasks.

Nevertheless, the aforementioned approaches primarily rely on explicit discrete modeling of CSI, in which channel acquisition is formulated as the direct recovery of CSI values at discrete grid points, leading to the following four limitations.
\begin{itemize}
    \item Grid-based discrete models directly reconstruct sampled CSI values without explicitly exploiting a frequency-aware inductive bias for multipath-induced high-frequency variations, which may limit their effectiveness in channel reconstruction tasks.
    \item High-dimensional features hinder accurate CSI feedback at low compression ratios. 
    Specifically, assume that the patch size along the time, frequency, and space dimensions is $(p_{\rm t}, p_{\rm f}, p_{\rm s})$, and the masking ratio is $1-r_{\rm m}$. 
    The input is first replicated $p_{\rm t}$ times along the time dimension to conform to the 3D CSI structure. 
    The compression ratio is defined as the ratio between the feedback tensor dimension and the original CSI dimension, i.e., $r_{\rm c} = \left(D_{\rm F} r_{\rm m} \frac{NKp_{\rm t}}{p_{\rm t} p_{\rm f} p_{\rm s}}\right) /  2NK = \frac{r_{\rm m}D_{\rm F}}{2p_{\rm f} p_{\rm s}}$. 
    For a typical setting with $D_{\rm F}=384$, $r_{\rm m}=1/10$, and $p_{\rm f}=p_{\rm s}=4$, the resulting compression ratio is $r_{\rm c}=1.2$, which is far from adequate for compressed feedback.
    \item As the number of CSI feature tokens, $N_{\rm F}$, scales with the input size, downstream fine-tuning is restricted to a fixed CSI size.
    When the CSI size changes, task-specific output heads must be redesigned and retrained, thereby increasing deployment cost.
    \item Existing approaches commonly adopt Transformer-based decoders, which introduce substantial inference latency and hinder deployment on resource-constrained wireless devices.
\end{itemize}

\section{INR-Based Unified Wireless Task Design}
In this section, we first present the motivation and network implementation of INR-based CSI representations. 
We then formulate a generalizable INR-based unified wireless system design and discuss its advantages.
\subsection{Motivation: Representing CSI as an INR}
As shown in \eqref{H-tkn} and \eqref{h-tf}, a CSI tensor essentially consists of samples from a channel function formed by multipath superposition. Therefore, CSI can be accurately represented using only a small number of multipath parameters, while the CSI value at any time-frequency-space coordinate can be queried with minimal computational cost. This parameterized representation, rather than a discrete tensor, enables compact and efficient CSI modeling. 
Nevertheless, accurately estimating multipath parameters in \eqref{h-tf} from partial and coarse CSI observations is highly challenging and computationally expensive, making it unsuitable for real-time channel acquisition. We therefore use a coordinate-conditioned INR to fit the sampled CSI without explicitly estimating the multipath parameters. 

Unlike explicit discrete modeling, INR represents CSI implicitly as a neural-network-based function $f_\theta$, which maps time-frequency-space coordinates to the corresponding complex CSI values, where $\theta$ denotes the neural network parameters.
Specifically, for any given three-dimensional coordinate $(s,k,n)$, we have $f_\theta(s,k,n)= \bm{H}_{s,k,n}$. 
Therefore, CSI is no longer treated as a discrete 3D tensor, but is instead modeled as a queryable neural function. 
In this case, all CSI tensors can be represented by a unified neural function $f_\theta$ with instance-specific compact parameters $\theta$.
In this paper, we adopt a SIREN-based neural network as the INR of CSI.
SIREN \cite{sitzmann2020implicit} is a representative INR architecture built upon an MLP, with conventional activation functions replaced by sinusoidal activations to map input coordinates to signal values.
Given the input $\bm{x}\in\mathbb{R}^{d_{\rm in}}$ and the output $\bm{y}\in\mathbb{R}^{d_{\rm out}}$, the single-layer form can be expressed as
\begin{align}
    \bm{y}= sin(\omega(\bm{W}\bm{x}+\bm{b})),
\end{align}
where $\bm{W}\in\mathbb{R}^{d_{\rm out}\times d_{\rm in}}$ and $\bm{b}\in\mathbb{R}^{d_{\rm out}}$ denote the weight matrix and bias vector, respectively, and $\omega$ is a hyperparameter that controls the frequency scale of the sinusoidal activation. 
By nonlinearly combining multiple sinusoidal basis functions, this architecture shares a certain similarity with the physical process by which CSI is formed through the superposition of multipath components. 
Therefore, SIREN is well-suited \cite{xiao2022cgrbfnet} for characterizing the oscillatory variations of CSI across the time, frequency, and space dimensions.
\subsection{Generalizable INR-based Unified Wireless System Design}
Despite their compact and efficient CSI representations, existing INR-based wireless methods typically require online optimization of a dedicated parameter set for each CSI instance. 
This per-instance optimization incurs substantial computational overhead, limiting its applicability to real-time tasks such as channel estimation and prediction. 
Inspired by recent advances in generalizable INRs \cite{kim2023generalizable}, we propose a generalizable INR architecture that unifies diverse wireless tasks while eliminating online parameter optimization for each CSI instance.

The proposed framework consists of two key components, i.e., an encoder $f_{enc}$ and a decoder $f_{\psi, \phi}$. 
The decoder acts as the INR of CSI, which maps each time-frequency-space coordinate $(s,k,n)$ to its corresponding CSI value $\bm{\hat H}_{s,k,n}$. 
The decoder parameters are divided into instance-specific parameters $\phi$ and instance-agnostic parameters $\psi$. 
The instance-agnostic parameters are shared across different CSI instances, while the instance-specific parameters are generated by a Transformer-based encoder conditioned on the current CSI instance $\bm{\tilde H}_{S}$. 
These instance-specific parameters are then used to modulate the decoder, enabling the INR to adapt to different CSI instances without online optimization. The overall processing pipeline is formulated as 
\begin{align}\label{equ: Genralizable}
\bm{\hat H}_{s,k,n} = f_{\psi, \phi}(s,k,n),\phi = f_{enc}(\bm{\tilde H}_{S}). 
\end{align}
Under this framework, the instance-agnostic parameters of the decoder can be viewed as a universal wireless basis, while the encoder can be regarded as a modulator of this basis. 

Unlike existing generalizable INRs in computer vision, which primarily focus on image compression, an INR-based wireless foundation model must accommodate heterogeneous CSI dimensions and incorporate diverse wireless-task-oriented pretraining strategies to generalize across scenarios, system configurations, and tasks.
Specifically, with masked reconstruction and interpolation-denoising pretraining tasks, it needs to support channel estimation and channel prediction. 
Moreover, a compression-enhanced pretraining strategy should be designed to further improve compression performance and learn transferable features that can be fine-tuned for various downstream wireless tasks. 
Compared with existing explicit discrete modeling-based wireless foundation models, the proposed framework offers the following prominent advantages.
\begin{itemize}
\item By treating CSI tensors as samples of multipath channel responses, the sinusoidal coordinate decoder introduces a frequency-aware inductive bias for fitting complex high-frequency variations.
This makes the proposed design well suited to channel reconstruction tasks.
\item In contrast to explicitly modeled feature tokens, the encoder-generated instance-specific parameters are highly compact, which substantially reduces the compression ratio for CSI feedback.
\item Since the encoder-generated instance-specific parameters are independent of the CSI size, they serve as size-generalizable features that can be fine-tuned for downstream tasks across different CSI dimensions.
\item The SIREN-based decoder incurs much lower inference latency than Transformer-based decoders, making the proposed framework more suitable for deployment on resource-constrained wireless devices.
\end{itemize}
A comparative summary of the two wireless foundation model design paradigms is presented in Fig. \ref{paradigm comparison}.
\begin{figure*}[!t]
    \centering
    \includegraphics[width=0.95\linewidth]{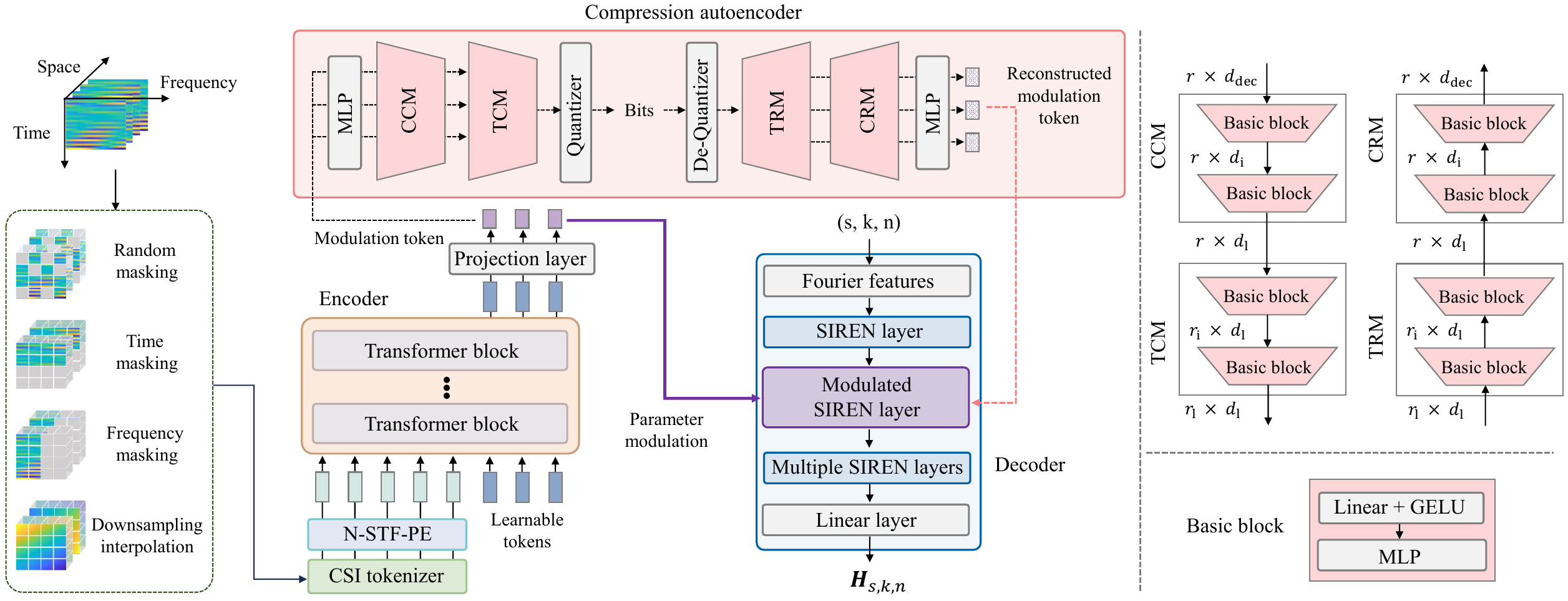}
    \vspace{-2mm}
    \caption{An illustration of the network structure of WiFo-INR.}
    \vspace{-4mm}
    \label{fig-network}
\end{figure*}
\section{Proposed WiFo-INR}
In this section, we present the proposed INR-based wireless foundation model, including its network architecture, pretraining strategy, and applications to CSI-related downstream tasks.
\subsection{Network Structure}
As an implementation of the framework given in \eqref{equ: Genralizable}, we design an INR-based wireless foundation model, termed WiFo-INR, whose network architecture is illustrated in Fig. \ref{fig-network}. WiFo-INR mainly consists of four components, namely a preprocessor for CSI preprocessing and tokenization, an encoder for generating modulation tokens, a decoder serving as the INR, and a compression autoencoder for reducing the compression ratio. These components are detailed below.
\subsubsection{Preprocessor}
It preprocesses the CSI and maps it into tokens. 
To facilitate neural network processing, the complex-valued CSI $\bm{H}$ is first converted into a real-valued representation $\bm{\tilde H}\in\mathbb{R}^{2\times T \times K \times N}$. 
We then divide $\bm{\tilde H}$ into non-overlapping 3D patches with patch size of $(p_{\rm t},p_{\rm f},p_{\rm s})$.
This process generates $L=\left\lceil \frac{T}{p_t} \right\rceil \left\lceil \frac{K}{p_f} \right\rceil \left\lceil \frac{N}{p_s} \right\rceil$ 3D patches, which are mapped into CSI tokens $\bm{X}_{\rm e} \in \mathbb{R}^{L\times d_{\rm enc}}$ through a 3D convolutional network.
Here $\lceil\cdot\rceil$ is the ceiling operator and $d_{\rm enc}$ denotes the embedding dimension of the encoder. 
For masked reconstruction pretraining tasks, only the visible CSI tokens $\bm{x}_{\rm vis} \in \mathbb{R}^{L_{\rm vis}\times d_{\rm enc}}$ are retained. 
For the interpolation-denoising pretraining task, all tokens of the coarsely estimated CSI are retained.
\subsubsection{Encoder}
It aims to extract instance-specific parameters from CSI tokens to modulate parameters in the decoder. 
The CSI tokens are first added with the normalized space-time-frequency absolute positional encoding (N-STF-PE) $\bm{P}_{\rm enc}\in \mathbb{R}^{L_{\rm vis}\times d_{\rm enc}}$, which differs from STF-PE \cite{liu2025wifo, liu2025WiFo2} only in that its input coordinates are normalized. 
This normalization aligns the input coordinates of the encoder and subsequent decoder, thereby eliminating the impact of varying CSI sizes.
Taking the time domain as an example, for the $i$-th token with temporal coordinate $p_i^{\rm t}$, the normalized coordinate is given by ${p_i^{\rm t}}/\max{((\left\lceil \frac{T}{p_t} \right\rceil-1),1)}\in [0,1]$. 
The position-encoded CSI tokens, together with $r$ learnable tokens $\bm{X}_{\rm l}\in \mathbb{R}^{r\times d_{\rm enc}}$, are then fed into a backbone built upon Transformer blocks with $d_{\rm enc}$ embedding dimension. 
The $r$ output tokens $\bm{X}_{\rm o}\in \mathbb{R}^{r \times d_{\rm enc}}$ corresponding to the learnable tokens are further passed through an MLP-based projection layer to obtain the modulation tokens $\bm{X}_{\rm m}\in \mathbb{R}^{r \times d_{\rm dec}}$, where $d_{\rm dec}$ represents the hidden width of the SIREN layer of the decoder.
\subsubsection{Decoder}
It aims to construct an INR that outputs the complex-valued CSI based on the time-frequency-space coordinates.
To enhance its capability of characterizing high-frequency variations across the time, space, and frequency dimensions, we transform the normalized input coordinate $\bm{c}=[\frac{s}{T-1}, \frac{k}{K-1},\frac{n}{N-1}]$ into Fourier features \cite{tancik2020fourier}, as follows. 
\begin{align}\label{fourier}
\gamma(\bm{c})=[\bm{c}, \sin(2\pi \bm{c} \bm{B}), \cos(2\pi \bm{c} \bm{B})]
\end{align}
Here, $\bm{B}\in \mathbb{R}^{3 \times d_{\rm f}}$ denotes the Gaussian matrix, and $d_{\rm f}$ is the number of frequency bases. The entries of $\bm{B}$ follow Gaussian distributions with mean 0 and domain-specific variances $\sigma_t^2$, $\sigma_s^2$, and $\sigma_f^2$ for the time, space, and frequency domains, respectively. 
The Fourier features are then transformed into low-level frequency patterns $\bm{h}_{\rm f}\in \mathbb{R}^{d_{\rm dec}}$ through a SIREN layer, as given by
\begin{align}
\bm{h}_{\rm f} = \sin\left(\omega_{\rm f}(\bm{W}_{\rm f} \gamma(\bm{c})+\bm{b}_{\rm f})\right),
\end{align}
where $\omega_{\rm f}$ denotes the frequency factor of the sinusoidal activation, and $\bm{W}_{\rm f}$ and $\bm{b}_{\rm f}$ denote the weight matrix and bias vector of the SIREN layer, respectively.
Subsequently, $\bm{h}_{\rm f}$ is processed by an $n_{\rm dec}$-layer SIREN network.
Following \cite{kim2023generalizable}, we use $\bm{X}_{\rm m}$ to modulate only the first SIREN layer, enabling more effective modulation of wireless patterns.
Specifically, let the instance-agnostic parameter of this SIREN layer be $\bm{U}\in\mathbb{R}^{d_{\rm dec}\times r}$, and the processing of this layer can then be expressed as
\begin{align}
\bm{h}_0=\sin\left(\omega_0(\bm{U}\bm{X}_{\rm m} \bm{h}_{\rm f}+\bm{b}_0)\right),
\end{align}
where $\bm{h}_0\in \mathbb{R}^{d_{\rm dec}}$ denotes the output hidden feature, $\omega_0$ is the frequency factor of the sinusoidal activation, and $\bm{b}_0$ is the bias.
The features are processed by an $(n_{\rm dec}-1)$-layer SIREN with frequency factor $\omega_0$ and then projected through a linear layer to obtain $\bm{h}_{\rm out}\in \mathbb{R}^{2}$, which is finally converted into the complex-valued CSI at the original coordinate $(s,k,n)$.
\subsubsection{Compression Autoencoder}
It is employed further to reduce the feedback overhead of the modulation tokens, enabling the model to support CSI compression feedback tasks with extremely high compression.
Its input is the modulation token $\bm{X}_{\rm m}$ generated by the encoder, and its output is the reconstructed modulation token $\bm{\hat X}_{\rm m}\in \mathbb{R}^{r \times d_{\rm dec}}$.
Specifically, the input is first processed by an MLP layer without changing its dimension, followed by a channel compression module (CCM) and a token compression module (TCM) that compress the channel dimension and the number of tokens, respectively. 
To enhance compression performance, each compression module adopts a two-stage progressive compression mechanism, where the features are first compressed to an intermediate scale and then to a lower scale. 
In each compression stage, a linear layer is first used to transform the dimensionality, followed by a cascaded MLP layer for feature refinement. 
Let $d_{\rm i}$ and $d_{\rm l}$ denote the channel dimensions of the two compression stages, and let $r_{\rm i}$ and $r_{\rm l}$ denote the corresponding numbers of tokens. 
The dimensionality changes during compression and reconstruction are illustrated in the right panel of Fig. \ref{fig-network}.
In this case, the compression ratio is derived as $r_{\rm c}=\frac{r_{\rm l}d_{\rm l}}{2NK}$.
The compressed modulation tokens are then quantized and dequantized using a non-uniform $\mu$-law quantizer with low-bit quantization \cite{liu2026wifo}. 
Finally, a symmetric network compressing a token reconstruction module (TRM) and a channel reconstruction module (CRM) is adopted to successively expand the token and channel dimensions, thereby reconstructing $\bm{\hat X}_{\rm m}$ for parameter modulation in the decoder.
\subsection{Pretraining Scheme}
\begin{figure}[!t]
    \centering
    \includegraphics[width=1\linewidth]{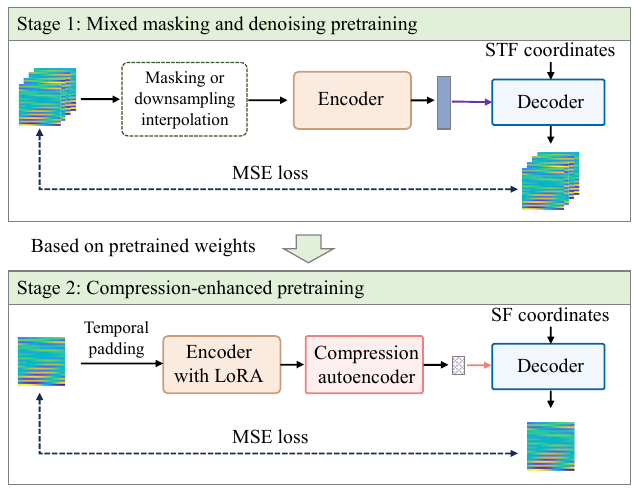}
    \vspace{-7mm}
    \caption{An illustration of pretraining schemes of WiFo-INR.}
    \vspace{-5mm}
    \label{fig-pretraining}
\end{figure}
WiFo-INR adopts a two-stage self-supervised pretraining scheme. 
The first stage employs mixed-masking and denoising objectives to learn generalizable 3D CSI representations, while the second stage uses compression-enhanced pretraining to adapt the model to 2D CSI feedback by reducing the dimensionality of the modulation tokens.
\subsubsection{Mixed Masking and Denoising Pretraining}
The first stage is designed for 3D CSI reconstruction, aiming to enhance the unified modeling capability of the model for channel reconstruction. 
During pretraining, the modulation tokens generated by the encoder directly modulate the decoder, while the compression autoencoder is deactivated. 
We consider the four pretraining tasks adopted in WiFo-2 \cite{liu2025WiFo2}. 
For random-masked reconstruction, all tokens are randomly masked with a high masking ratio to strengthen the general time-frequency-space modeling capability of the model. 
For time-masked and frequency-masked reconstruction, tokens located in the last fraction of the time or frequency domain are masked, respectively, which encourages the model to learn predictive representations for time-domain and frequency-domain channel prediction. 
For interpolation-denoising, the original CSI is first downsampled at a given ratio to emulate the partial pilot observations in channel estimation. 
Linear interpolation is then applied to obtain a coarse CSI estimate with the same size as the original CSI, which is fed into the network. 
This task encourages the model to recover accurate CSI from coarse estimates, thereby improving its channel estimation capability. 
In this stage, one pretraining strategy is randomly selected for each batch, and the mean squared error (MSE) is adopted as the loss function:
\begin{align}
\mathcal{L}{\rm rec}=\frac{\|\bm{H}-\bm{H}_{\rm rec}\|_F^2}{|\bm{H}|},
\end{align}
where $\bm{H}_{\rm rec}$ denotes the reconstructed CSI and $|\bm{H}|$ denotes the number of elements in $\bm{H}$.
\subsubsection{Compression-Enhanced Pretraining}
The second stage is designed for 2D CSI compression feedback, where the modulation tokens generated by the encoder are dimension-reduced and quantized to decrease the compression ratio further.
To enable efficient training from the first-stage pretrained weights, the encoder is adapted using LoRA, with all pretrained encoder parameters frozen except for the patch embedding layer, while all other modules remain trainable.
Since CSI feedback operates on the two-dimensional frequency-space CSI $\bm{H}_{\rm CF}\in\mathbb{C}^{K\times N}$, we first replicate it $p_{\rm t}$ times along the temporal dimension to form $\bm{\widetilde{H}}_{\rm CF}$, thereby matching the network input dimensions. 
This operation preserves the knowledge learned from three-dimensional CSI during the first stage, as such padding can be viewed as a special case of temporally quasi-static channels.
In this case, $\bm{\tilde H}_{\rm CF}$ is tokenized directly and fed into the encoder without masking or other preprocessing. 
The decoder is then modulated by the reconstructed modulation tokens $\bm{\hat X}_{\rm m}$ from the compression autoencoder.
Finally, the decoder outputs the CSI at time coordinate 0 as the reconstructed CSI. 
As in the first stage, we use MSE as the loss function.

\subsection{Applications}
After pretraining, two sets of network weights are obtained and deployed for various wireless tasks.
\subsubsection{Channel Reconstruction}
    WiFo-INR can be deployed at either the BS or UE side for channel reconstruction tasks. 
    In this case, the complete model except the compression autoencoder is deployed, and the pretrained weights from the first stage are loaded. 
    For time-domain and frequency-domain channel prediction, the partial CSI in the time or frequency domain is treated as the visible input to reconstruct the complete CSI. 
    For channel estimation, the coarse CSI estimate obtained via linear interpolation is fed into the network to recover an accurate CSI estimate.
\subsubsection{CSI Feedback}
    In this case, the encoder and the compression module of the compression autoencoder are deployed at the UE, while the remaining network modules are deployed at the BS. 
    The pretrained weights from the second stage are loaded.
    Similar to the second-stage pretraining process, the 2D CSI at the UE side is first padded along the time dimension and then fed into the network to generate compressed modulation tokens.
    The tokens are quantized and fed back to the BS over the air interface, where the modulation tokens are reconstructed from the received bits for parameter modulation. 
    Then the CSI is reconstructed according to the 2D coordinates.
\subsubsection{CSI-Related Tasks}
    WiFo-INR can also be applied to CSI-related tasks under either single-sided or two-sided deployment. 
    The original or compressed modulation tokens are used as CSI features and fed into lightweight output heads for downstream tasks. 
    During the fine-tuning process, only the output heads are trainable.
    In this paper, we consider the following three representative tasks.
\begin{itemize}
    \item \textbf{Wireless localization}: The UE estimates its position from downlink 2D CSI. 
    The second-stage pretrained weights are loaded. 
    The UE maps the 2D CSI into compressed modulation tokens and employs a lightweight regression head to predict the 2D coordinates.
    \item \textbf{Beam prediction}: The UE compresses and feeds back the downlink sub-6 GHz CSI to the BS, which predicts the optimal mmWave beam index. 
    The second-stage pretrained weights are loaded. 
    The UE compresses the 2D CSI into bits for feedback, while the BS recovers the compressed modulation tokens from the received bits and employs a classification head to predict the beam index.
    \item \textbf{Scenario classification}: The UE classifies the propagation scenario as LoS or NLoS based on 3D CSI. 
    The first-stage pretrained weights are loaded. 
    The UE maps the 3D CSI into modulation tokens and employs a classification head to predict the scenario class.
\end{itemize}
\begin{table*}[!htbp]
\footnotesize
\caption{An illustration of the system configurations of the constructed large-scale heterogeneous CSI dataset. $f_{\rm c}$ denotes the center frequency.}
\vspace{-2mm}
\label{dataset}
\centering
\begin{tabular}{ccccccccccc}
\toprule[1pt]
Usage & Source & Dataset & \makecell{$f_{\rm c}$\\(GHz)} & $K$ & \makecell{$\Delta f$\\(kHz)} & T & \makecell{$\Delta t$\\(ms)} & Antenna & Scenario & \makecell{User speed\\(km/h)} \\ \midrule[1pt]
\multirow{2}{*}{Pretraining}& \multirow{2}{*}{QuaDRiGa}   &PC1-PC16  &  \multirow{2}{*}{1.5-5.9}   &  32-128 & 16-32 & 60-360  & 0.5-1    & \multirow{2}{*}{\makecell{(1-4) \\$\times$ (4-8)}}    & \multirow{2}{*}{\makecell{UMi/UMa/RMa/\\ Indoor +  LoS/NLoS}} &  \multirow{2}{*}{0-200}    \\ 
    &&PF1-PF16 &     & 36-72 & 15-60 & 12-28 & 0.017-0.067 &     &      &    \\\hline
\multirow{8}{*}{\makecell{Generalization\\validation}}&\multirow{6}{*}{QuaDRiGa}&GC1 &  \multirow{2}{*}{6.7}   & 64 & 60 & 16 & 1 & \multirow{2}{*}{$4\times4$}  & \multirow{2}{*}{UMi LoS}     &  \multirow{2}{*}{3-50}  \\
&&GF1 &     & 36 & 15 & 12 & 0.067 &   &      &   \\\cmidrule(lr){3-11}
&&GC2 &  \multirow{2}{*}{28}   & 64 & 90 & 16 & 0.5 & \multirow{2}{*}{$2\times4$}  & \multirow{2}{*}{InF NLoS}     &  \multirow{2}{*}{0-5}  \\
&&GF2 &     & 72 & 30 & 28 & 0.033 &  &     &  \\\cmidrule(lr){3-11}
&&GC3 &  \multirow{2}{*}{40}   & 64 & 360 & 16 & 0.5 & \multirow{2}{*}{$4\times4$}  & \multirow{2}{*}{Indoor NLoS}     &  \multirow{2}{*}{0-10}  \\
&&GF3 &     & 36 & 30 & 12 & 0.033 &  &     &  \\\cmidrule(lr){2-11}
&Sionna RT&GC4 &  5.9   & 64 & 180 & 16 & 2 & $1\times4$  & Pedestrian     &  5  \\ \cmidrule(lr){2-11}
&DeepMIMO&GF4 &  200   & 72 & 120 & 12 & 0.005 & $2\times8$  & Drone     &  30-50  \\
\bottomrule[1pt]   
\end{tabular}
\vspace{-5mm}
\end{table*}
\section{Experimental Results}
In this section, we first introduce the dataset construction and simulation setup. We then evaluate the performance on various wireless tasks, followed by ablation studies and scaling analysis.
\subsection{Dataset Construction}
\subsubsection{Channel Recovery}
To support the two-stage pretraining and generalization validation, we construct a large-scale and heterogeneous CSI dataset. 
It consists of 32 and 8 sub-datasets for pretraining and generalization validation, respectively, categorized as either coarse-grained or fine-grained. 
The fine-grained CSI adopts an OFDM-symbol-level time-domain sampling interval and a small subcarrier spacing to match the settings of channel estimation, while the coarse-grained CSI adopts slot-level temporal sampling and a larger subcarrier spacing to meet the requirements of channel prediction and CSI feedback. 
Each sub-dataset contains 12,000 time-frequency-space CSI samples, of which 9,000, 1,000, and 2,000 samples are used for training, validation, and testing, respectively.
For the pretraining set, we generate 16 coarse-grained CSI datasets (PC1–PC16) and 16 fine-grained CSI datasets (PF1–PF16) using QuaDRiGa \cite{jaeckel2014quadriga}. 
To enable WiFo-INR to learn general wireless representations, the pretraining set spans carrier frequencies from 1.5 GHz to 5.9 GHz, UMa, UMi, RMa, and Indoor scenarios under both LoS and NLoS conditions, user speeds from 0 to 200 km/h, and diverse antenna, subcarrier, and temporal sampling configurations.
The generalization validation set contains datasets from three sources, namely QuaDRiGa, Sionna RT \cite{aitaoudia2025sionna}, and DeepMIMO \cite{alkhateeb2019deepmimo}. 
Specifically, we generate 3 coarse-grained datasets (GC1–GC3) and three fine-grained datasets (GF1–GF3) using QuaDRiGa. They cover three unseen carrier frequencies of 6.7 GHz, 28 GHz, and 40 GHz, as well as the unseen indoor factory (InF)-NLoS scenario, to evaluate the generalization capability across frequencies and scenarios. 
Building on Sionna RT, we construct a coarse-grained dataset (GC4) for low-speed pedestrian mobility in the Peking University campus scenario, as detailed in \cite{liu2025WiFo2}.
In addition, we generate a fine-grained dataset (GF4) based on the drone scenario in DeepMIMO, with an unseen carrier frequency of 200 GHz.
The detailed system parameters of each sub-dataset are provided in Table \ref{dataset}.

\subsubsection{CSI-Related Downstream Tasks}
(i) For wireless localization, we adopt the real-world DICHASUS measurement dataset and construct 1,500 samples from the dichasus-005x scenario, including 1,000 and 500 samples for training and testing, respectively. 
Each sample consists of 2D space-frequency CSI and its corresponding 2D position. 
The CSI is measured at 1.272 GHz with 64 subcarriers and a $4\times8$ UPA.
(ii) For beam prediction, we generate 2,000 samples using QuaDRiGa under the UMi-LoS scenario, with 1,000 and 1,000 samples used for training and testing, respectively. 
Each sample contains space–frequency 2D CSI collected at 2.5 GHz and the optimal beam at 28 GHz selected from a 64-entry DFT codebook, both obtained at the same transmitter–receiver locations. 
The sub-6 GHz link covers 64 subcarriers and employs a $1\times16$ ULA, while the mmWave link employs a $1\times64$ ULA.
(iii) For scenario classification, we generate data using QuaDRiGa under the UMi-LoS and UMi-NLoS scenarios at 2.5 GHz. 
To evaluate generalization across different CSI dimensions, 6 sub-datasets are used for joint training, while one additional sub-dataset is reserved for generalization testing. 
Each sub-dataset contains 600 samples, including 100 and 500 samples for training and testing, respectively, with equal proportions of LoS and NLoS samples. 
\subsection{Simulation Setup}
\subsubsection{Network and Pretraining Settings}
\begin{table}[!t]
\footnotesize
\caption{Model configurations of different WiFo-INR variants.}
\vspace{-2mm}
\label{model-version}
\centering
\setlength{\tabcolsep}{3.5pt}
\begin{tabular}{cccccc}
\toprule[1pt]
Version & Enc. width & Enc. heads & Dec. width & Dec. depth & Params. (M)  \\
\midrule
Tiny   & 96  & 2 & 64  & 3 & 0.74  \\
Little & 138 & 3 & 88  & 3 & 1.50  \\
Small  & 192 & 4 & 128 & 4 & 2.90  \\
Base   & 270 & 6 & 176 & 5 & 5.72  \\
Large  & 384 & 8 & 256 & 6 & 11.59 \\
\bottomrule[1pt]
\end{tabular}
\vspace{-5mm}
\end{table}
To accommodate wireless devices under different system constraints, we consider five WiFo-INR variants, namely Tiny,  Little, Small, Base, and Large. 
Their detailed configurations are given in Table \ref{model-version}, where decoder depth denotes the number of SIREN layers $n_{\rm dec}$ after the low-level frequency patterns. 
For all variants, the patch size in the time, frequency, and space domains is fixed to (4,4,4). 
The encoder contains 6 Transformer blocks with an MLP ratio of 4 and is equipped with $r=32$ modulation tokens. 
For the compression autoencoder, the two-stage compression dimensions and token numbers are set to $d_{\rm i}=64$, $d_{\rm l}=16$, $r_{\rm i}=16$, and $r_{\rm l}=4$, respectively. 
In the decoder, $\sigma_t^2$, $\sigma_s^2$, and $\sigma_f^2$ are all set to 1 for Gaussian-sampled frequency bases. 
The frequency factors of the sinusoidal activations are set to $\omega_{\rm f}=10$ and $\omega_0=1$.
During both pretraining and inference, CSI is first subjected to per-sample power normalization.

For the first pretraining stage, all 32 pretraining datasets are used. 
Specifically, PC1-PC16 are used for the masked reconstruction pretraining tasks, while PF1-PF16 are used for the interpolation-denoising pretraining task. 
For the second pretraining stage, PC1-PC16 are adopted, and only the 2D CSI slice at the first time instant of each CSI sample is used.
During the first pretraining stage, the original CSI is randomly corrupted by Gaussian noise with SNRs ranging from 10 to 25 dB. For random-masked, time-masked, and frequency-masked reconstruction, the masking ratios are set to 80\%, 25\%, and 25\%, respectively. 
For interpolation-denoising pretraining, the pilot placement ratios in the time, frequency, and space domains are set to 1/4, 1/12, and 1, respectively. 
During the second pretraining stage, the number of quantization bits is fixed at 5, and the LoRA rank and scaling factor in the encoder are set to 8 and 16, respectively.
Both pretraining stages use a batch size of 128 and are trained for 1000 epochs with AdamW ($\beta_1=0.9$, $\beta_2=0.999$) with a weight decay of $1\times 10^{-4}$. 
We adopt a cosine-decay learning-rate schedule with 5 warmup epochs. 
The base and minimum learning rates are $3\times 10^{-4}$ and $1\times 10^{-7}$ in the first stage, and $1\times 10^{-4}$ and $1\times 10^{-7}$ in the second stage, respectively. 
All models are trained with FP32 precision on 4 NVIDIA GeForce RTX 4090 GPUs.
\subsubsection{Baselines}
To comprehensively compare WiFo-INR with existing methods across diverse wireless tasks, we consider the following four categories of representative baselines. 
For each AI baseline, we selected the most appropriate training hyperparameters and trained it sufficiently to ensure its best attainable performance.
\begin{itemize}
\item \textbf{Wireless foundation models}: 
We adopt WiFo-2 \cite{liu2025WiFo2}, a state-of-the-art (SOTA) wireless foundation model based on explicit discrete modeling, as a baseline for channel reconstruction and CSI-related downstream tasks. 
For a fair comparison, we use its dense variant, excluding the MoE architecture, and pretrain it with the same data and protocol as the first-stage pretraining of WiFo-INR. 
To approximately match the parameter count, it employs a six-layer encoder and a one-layer decoder, with the encoder depth and hidden dimensions aligned with WiFo-INR-Large. 
Since WiFo-2 is not designed for CSI feedback at low compression ratios, we further include the base version of WiFo-CF \cite{liu2026wifo} as a dedicated baseline, following its original configuration and pretraining scheme while using the same data as the second-stage pretraining of WiFo-INR.
\item \textbf{Task-specific AI models}: 
Task-specific AI models are designed for particular wireless tasks and are fully trained and evaluated on the specific dataset.
For both the time domain and frequency domain channel prediction tasks, we adopt the Transformer \cite{jiang2022accurate} as the task-specific baseline.
For channel estimation, we utilize Channelformer \cite{luan2023channelformer}, where the antenna dimension is processed in parallel as part of the batch dimension. 
For CSI feedback, the autoencoder-based CsiNet \cite{wen2018deep} and Transformer-based TransNet \cite{cui2022transnet} are adopted as representative AI-based schemes. 
For the three CSI-related downstream tasks, we employ BP-DNN \cite{alrabeiah2020deep}, WiT \cite{salihu2022attention}, and ST-CNN \cite{sun2022channel} for beam prediction, wireless localization, and scenario classification, respectively.

\item \textbf{LLM-based schemes}: 
LLM-based schemes \cite{liu2025llm4wm} refer to methods that enable cross-domain knowledge transfer by fine-tuning LLMs on the specific dataset.
For channel prediction tasks, we consider LLM4CP \cite{liu2024llm4cp}, which fine-tunes a pretrained GPT-2 model for CSI prediction. 
Similarly, the same architecture is adapted to channel estimation by formulating the task as CSI denoising, termed LLM4CE.

\item \textbf{Interpolation/extrapolation}: 
The first-order linear extrapolation is adopted for channel prediction, while the bilinear interpolation is used for channel estimation.
\end{itemize}
\subsubsection{Evaluation Metrics}
We comprehensively evaluate the models from two aspects: task performance and inference cost.
\begin{itemize}
\item \textbf{Task performance}: 
For channel prediction tasks, we measure the prediction accuracy using the NMSE between the predicted CSI entries and the ground truth. 
For channel estimation and CSI feedback, NMSE is also adopted, but it is computed over the complete CSI tensor. 
For beam prediction, wireless localization, and scenario classification, we use the top-1 classification accuracy, the root-mean-square error (RMSE) between the predicted and ground-truth 2D user coordinates, and the F1 score as the performance metrics, respectively.

\item \textbf{Inference cost}: 
We consider three metrics to evaluate the deployment feasibility on communication devices with storage, power, and latency constraints. 
Specifically, the number of parameters measures the storage overhead, floating point operations (FLOPs) measure the theoretical complexity of model inference, and inference time measures the latency overhead incurred by model inference.
\end{itemize}
\begin{table}[!t]
\footnotesize
\caption{Performance (NMSE in dB) comparison between WiFo-INR and other baselines on the time-domain channel prediction task. 
The best and second-best results are highlighted in \textbf{bold} and \underline{underlined}, respectively.}
\vspace{-2mm}
\label{result-CPT}
\centering
\setlength{\tabcolsep}{3pt} 
\begin{tabular}{c|ccccc}
\toprule[1pt]
Dataset & WiFo-INR & WiFo-2 & Transformer & LLM4CP & Extrapolation \\
\midrule
PC1  & \textbf{-14.97} & \underline{-13.73} & -9.31  & -12.55 & 3.88   \\
PC2  & \textbf{-10.58} & \underline{-9.07}  & -5.28  & -7.70  & 2.86   \\
PC3  & \textbf{-30.99} & \underline{-27.05} & -19.76 & -23.58 & -20.03 \\
PC4  & \underline{-15.02} & \textbf{-16.78} & -5.82  & -13.39 & 1.86   \\
PC5  & \textbf{-11.54} & \underline{-10.04} & -7.46  & -8.42  & 3.48   \\
PC6  & \textbf{-12.80} & \underline{-12.02} & -6.81  & -9.39  & 3.71   \\
PC7  & \underline{-11.99} & \textbf{-13.65} & -2.45  & -9.81  & 4.58   \\
PC8  & \textbf{-27.51} & \underline{-25.36} & -15.82 & -20.76 & -14.39 \\
PC9  & \textbf{-11.06} & \underline{-10.21} & -6.85  & -9.23  & 3.72   \\
PC10 & \underline{-9.23}  & \textbf{-9.92}  & -5.09  & -7.37  & 4.24   \\
PC11 & \underline{-9.36}  & \textbf{-9.87}  & -2.47  & -7.75  & 4.20   \\
PC12 & \textbf{-24.84} & \underline{-23.43} & -14.62 & -18.16 & -5.36  \\
PC13 & \underline{-7.22}  & \textbf{-7.96}  & -2.78  & -5.89  & 3.68   \\
PC14 & \textbf{-10.67} & \underline{-9.93}  & -5.70  & -8.08  & 3.38   \\
PC15 & \textbf{-24.58} & \underline{-22.77} & -15.11 & -15.20 & -3.18  \\
PC16 & \textbf{-8.46}  & \underline{-8.35}  & -1.75  & -6.52  & 3.61   \\
\hline
\rowcolor{gray!15}
\makecell{\textbf{Avg.}\\(PC1-PC16)}
      & \textbf{-15.05} & \underline{-14.38} & -7.94 & -11.49 & 0.01 \\
\midrule
GC1 & \textbf{-14.57} & \underline{-13.58} & -6.85  & -10.78 & 3.98   \\
GC2 & \textbf{-12.07} & \underline{-11.84} & -10.13 & -11.43 & 3.24   \\
GC3 & \textbf{-21.11} & \underline{-20.44} & -18.24 & -20.00 & -0.89  \\
GC4 & \textbf{-21.20} & \underline{-20.55} & -16.40 & -18.24 & -15.26 \\
\hline
\rowcolor{gray!15}
\makecell{\textbf{Avg.}\\(GC1-GC4)}
& \textbf{-17.24} & \underline{-16.60} & -12.91 & -15.11 & -2.23 \\
\bottomrule[1pt]
\end{tabular}
\vspace{-4mm}
\end{table}
\begin{table}[!t]
\footnotesize
\caption{Performance comparison (NMSE in dB) between WiFo-INR and other baselines on the frequency-domain channel prediction task. 
The best and second-best results are highlighted in \textbf{bold} and \underline{underlined}, respectively.}
\vspace{-2mm}
\label{result-CPF}
\centering
\setlength{\tabcolsep}{3pt} 
\begin{tabular}{c|ccccc}
\toprule[1pt]
Dataset & WiFo-INR & WiFo-2 & Transformer & LLM4CP & Extrapolation \\
\midrule
PC1  & \textbf{-8.59}  & -7.94  & -4.97  & \underline{-8.31}  & 2.50  \\
PC2  & \textbf{-12.29} & \underline{-10.44} & -1.72  & -8.35  & 0.82  \\
PC3  & \textbf{-26.93} & \underline{-22.65} & -16.21 & -19.58 & -6.85 \\
PC4  & \textbf{-12.50} & \underline{-11.26} & -4.09  & -7.96  & -0.91 \\
PC5  & \textbf{-16.81} & \underline{-14.65} & -7.88  & -11.93 & -3.46 \\
PC6  & \textbf{-13.49} & \underline{-11.48} & -4.92  & -9.87  & 0.21  \\
PC7  & \textbf{-10.51} & \underline{-9.45}  & -6.98  & -6.48  & -0.22 \\
PC8  & \textbf{-20.51} & \underline{-17.39} & -10.91 & -15.31 & 0.07  \\
PC9  & \textbf{-9.12}  & \underline{-8.12}  & -4.43  & -7.79  & 2.03  \\
PC10 & \textbf{-10.33} & \underline{-9.17}  & -3.50  & -6.88  & -0.35 \\
PC11 & \textbf{-8.40}  & \underline{-7.00}  & -4.52  & -5.08  & 2.44  \\
PC12 & \textbf{-25.25} & \underline{-22.09} & -12.34 & -16.40 & -6.88 \\
PC13 & \textbf{-6.99}  & \underline{-5.80}  & -3.52  & -4.18  & 2.27  \\
PC14 & \textbf{-10.02} & \underline{-8.63}  & -2.61  & -7.15  & 1.87  \\
PC15 & \textbf{-26.18} & \underline{-22.55} & -12.30 & -19.33 & -6.86 \\
PC16 & \textbf{-10.07} & \underline{-8.72}  & -6.79  & -7.13  & 1.95  \\
\hline
\rowcolor{gray!15}
\makecell{\textbf{Avg.}\\(PC1-PC16)}
     & \textbf{-14.25} & \underline{-12.33} & -6.73 & -10.11 & -0.71 \\
\midrule
GC1 & \textbf{-13.90} & \underline{-12.08} & -8.72  & -9.11  & 0.37   \\
GC2 & \textbf{-12.85} & \underline{-11.63} & -8.13  & -11.55 & 2.20   \\
GC3 & \textbf{-16.10} & -14.08 & -11.74 & \underline{-15.38} & 2.89   \\
GC4 & \textbf{-21.79} & -17.20 & -17.23 & \underline{-18.74} & -11.94 \\
\hline
\rowcolor{gray!15}
\makecell{\textbf{Avg.}\\(GC1-GC4)}
& \textbf{-16.16} & \underline{-13.75} & -11.45 & -13.69 & -1.62 \\
\bottomrule[1pt]
\end{tabular}
\vspace{-5mm}
\end{table}
\begin{table}[!t]
\footnotesize
\caption{Performance comparison (NMSE in dB) between WiFo-INR and other baselines on the channel estimation task. 
The best and second-best results are highlighted in \textbf{bold} and \underline{underlined}, respectively.}
\vspace{-2mm}
\label{result-CE}
\centering
\setlength{\tabcolsep}{3pt}
\begin{tabular}{c|ccccc}
\toprule[1pt]
Dataset & WiFo-INR & WiFo-2 & Channelformer & LLM4CE & Interpolation \\
\midrule
PF1  & \textbf{-26.44} & \underline{-24.61} & -19.41 & -18.64 & -15.30 \\
PF2  & \textbf{-24.86} & \underline{-23.83} & -18.27 & -16.57 & -15.12 \\
PF3  & \textbf{-31.87} & \underline{-28.46} & -24.25 & -20.63 & -17.78 \\
PF4  & \textbf{-12.83} & \underline{-12.53} & -10.42 & -10.22 & -8.40  \\
PF5  & \textbf{-21.83} & \underline{-21.16} & -16.58 & -15.95 & -13.82 \\
PF6  & \textbf{-29.42} & \underline{-26.86} & -20.38 & -19.23 & -16.98 \\
PF7  & \underline{-9.46}  & \textbf{-9.57} & -8.40  & -8.22  & -6.25  \\
PF8  & \textbf{-26.77} & \underline{-25.50} & -20.40 & -21.01 & -17.13 \\
PF9  & \textbf{-18.62} & \underline{-18.15} & -15.01 & -14.26 & -11.76 \\
PF10 & \underline{-13.69} & \textbf{-13.86} & -12.00 & -11.54 & -9.56  \\
PF11 & \textbf{-13.30} & \underline{-13.22} & -10.09 & -10.60 & -7.18  \\
PF12 & \textbf{-28.70} & \underline{-26.58} & -22.02 & -20.42 & -17.32 \\
PF13 & \textbf{-17.96} & \underline{-16.78} & -11.72 & -11.62 & -5.91  \\
PF14 & \underline{-19.77} & \textbf{-19.81} & -15.44 & -14.98 & -12.33 \\
PF15 & \textbf{-27.55} & \underline{-25.91} & -20.54 & -21.64 & -17.38 \\
PF16 & \underline{-13.64} & \textbf{-13.65} & -10.63 & -10.84 & -7.77  \\
\hline
\rowcolor{gray!15}
\makecell{\textbf{Avg.}\\(PF1-PF16)}
& \textbf{-21.04} & \underline{-20.03} & -15.97 & -15.40 & -12.50 \\
\midrule
GF1 & \textbf{-25.28} & \underline{-23.90} & -6.08  & -17.78 & -15.32 \\
GF2 & \textbf{-25.39} & \underline{-23.94} & -20.00 & -20.00 & -17.70 \\
GF3 & \textbf{-28.14} & \underline{-25.88} & -23.01 & -22.22 & -17.75 \\
GF4 & \textbf{-20.77} & -17.03 & -13.24 & \underline{-18.78} & -15.63 \\
\hline
\rowcolor{gray!15}
\makecell{\textbf{Avg.}\\(GF1-GF4)}
& \textbf{-24.89} & \underline{-22.69} & -15.58 & -19.69 & -16.60 \\
\bottomrule[1pt]
\end{tabular}
\vspace{-4mm}
\end{table}
\subsection{Performance Evaluation on Wireless Tasks}
\subsubsection{Channel Reconstruction}
We first evaluate the performance of WiFo-INR-Large on channel reconstruction tasks, including full-shot performance on the pretraining datasets and zero-shot performance on the generalization datasets. 
For time-domain and frequency-domain channel prediction, the prediction ratio is set to 25\%. 
For channel estimation, the pilot symbol placement ratios along the time, frequency, and space dimensions are set to (1/4, 1/12, 1). 
We consider imperfect CSI by adding Gaussian noise with the SNR fixed at 20 dB.
WiFo-2 is also evaluated via zero-shot inference on the generalization datasets, whereas the other AI baselines are trained and tested separately on each sub-dataset. 
The performance of all models on the three tasks is reported in Tables \ref{result-CPT}, \ref{result-CPF}, and \ref{result-CE}.
On the pretraining datasets, WiFo-INR achieves the lowest average NMSE across all three tasks, reducing NMSE by a further 1.20 dB compared with the second-best WiFo-2 scheme, and obtains the best or second-best performance on all datasets. 
This demonstrates that the proposed architecture enables effective joint training across heterogeneous CSI datasets.
On the generalization datasets, WiFo-INR surpasses WiFo-2 and achieves SOTA zero-shot performance. 
Notably, it substantially outperforms the full-shot performance of all competing methods, reducing the average NMSE across the three tasks by 3.27 dB. 
These results highlight the strong and generalizable channel reconstruction capability of WiFo-INR.

Table \ref{cost-cr} reports the inference costs of different schemes. We consider single-sample inference on the PC13 dataset for time-domain channel prediction with a prediction ratio of 25\%. 
WiFo-INR achieves the lowest inference time among all AI-based schemes. 
Compared with WiFo-2, its lightweight decoder reduces the number of parameters, FLOPs, and inference time by 20.94\%, 27.59\%, and 44.56\%, respectively. 
These results demonstrate that, compared with MAE-based models, INR-based wireless foundation models achieve superior performance with substantially lower inference costs, facilitating deployment on resource-constrained communication devices.
\begin{table}[!t]
\footnotesize
\caption{Inference cost comparison among different channel reconstruction schemes.}
\vspace{-2mm}
\label{cost-cr}
\centering
\setlength{\tabcolsep}{2.5pt}
\begin{tabular}{c|ccccc}
\toprule[1pt]
 & WiFo-INR & WiFo-2 & Transformer & LLM4CP & Extrapolation \\
\midrule
Params (M) & 11.59 & 14.66 & 1.08 & 82.84 & / \\
FLOPs (G) & 3.36 & 4.64 & 0.37 & 1.76 & $0.18\times 10^{-3}$ \\
\makecell[c]{Inference\\ time (ms)}
& 3.62 & 6.53 & 8.49 & 5.79 & 0.17 \\
\bottomrule[1pt]
\end{tabular}
\vspace{-5mm}
\end{table}
\begin{table}[!t]
\footnotesize
\caption{Performance comparison (NMSE in dB) between WiFo-INR and other baselines on the CSI feedback task. 
The best and second-best results are highlighted in \textbf{bold} and \underline{underlined}, respectively.}
\vspace{-2mm}
\label{result-csi-feedback}
\centering
\setlength{\tabcolsep}{5pt}
\begin{tabular}{c|c|cccc}
\toprule[1pt]
Dataset & CR &
WiFo-INR & WiFo-CF & CsiNet & TransNet \\
\midrule
PC1  & 1/16 & \textbf{-14.16} & /      & -8.93  & \underline{-11.61} \\
PC2  & 1/32 & \textbf{-16.04} & \underline{-13.28} & -9.03  & -11.67 \\
PC3  & 1/16 & \textbf{-23.05} & /      & \underline{-18.24} & -15.69 \\
PC4  & 1/32 & \textbf{-8.91}  & \underline{-6.52}  & -4.09  & -6.46  \\
PC5  & 1/8  & \textbf{-20.45} & /      & \underline{-15.69} & -14.20 \\
PC6  & 1/32 & \textbf{-14.88} & \underline{-11.94} & -7.85  & -11.08 \\
PC7  & 1/32 & \textbf{-7.46}  & \underline{-5.33}  & -3.37  & \underline{-5.33}  \\
PC8  & 1/32 & \textbf{-14.16} & \underline{-12.15} & -6.74  & -12.01 \\
PC9  & 1/16 & \textbf{-13.87} & /      & -8.89  & \underline{-10.92} \\
PC10 & 1/16 & \textbf{-9.81}  & /      & -5.47  & \underline{-7.90}  \\
PC11 & 1/32 & \textbf{-8.36}  & \underline{-6.62}  & -3.84  & -5.64  \\
PC12 & 1/32 & \textbf{-12.83} & -10.46 & -5.53  & \underline{-11.08} \\
PC13 & 1/32 & \textbf{-6.52}  & \underline{-4.47}  & -2.72  & -4.45  \\
PC14 & 1/32 & \textbf{-13.22} & \underline{-11.25} & -7.54  & -10.13 \\
PC15 & 1/32 & \textbf{-18.74} & \underline{-15.23} & -10.41 & -15.09 \\
PC16 & 1/32 & \textbf{-9.64}  & \underline{-6.42}  & -3.72  & -6.40  \\
\hline
\rowcolor{gray!15}
\makecell{\textbf{Avg.}\\(PC1-PC16)}
& -- & \textbf{-13.26} & / & -7.63 & \underline{-9.98} \\
\midrule
GC1 & 1/32 & \textbf{-14.01} & \underline{-11.94} & -8.86  & -11.37 \\
GC2 & 1/32 & \textbf{-10.74} & \underline{-8.79}  & -4.26  & -8.18  \\
GC3 & 1/16 & \textbf{-17.27} & /      & -9.87  & \underline{-13.01} \\
GC4 & 1/16 & \underline{-18.49} & /      & -17.96 & \textbf{-18.54} \\
\hline
\rowcolor{gray!15}
\makecell{\textbf{Avg.}\\(GC1-GC4)}
& -- & \textbf{-15.13} & / & -10.24 & \underline{-12.77} \\
\bottomrule[1pt]
\end{tabular}
\vspace{-4mm}
\end{table}
\begin{table}[!t]
\footnotesize
\caption{Inference cost comparison among different CSI feedback schemes, with values before and after the slash denoting the UE-side and BS-side costs, respectively.}
\label{cost-cf}
\centering
\setlength{\tabcolsep}{5pt}
\begin{tabular}{c|cccc}
\toprule[1pt]
 & WiFo-INR & WiFo-CF & CsiNet & TransNet \\
\midrule
Params (M) & 11.84/0.69 & 1.91/3.65 & 0.065/0.070 & 0.41/0.43 \\
FLOPs (G) & 2.24/3.25 & 0.40/0.74 & \makecell{(0.11/3.5)\\$\times 10^{-3}$} & \makecell{(16.9/16.9)\\$\times 10^{-3}$} \\
\makecell[c]{Inference\\ time (ms)}
& 7.65/1.78 & 8.94/14.44 & 0.14/0.60 & 0.92/1.59 \\
\bottomrule[1pt]
\end{tabular}
\vspace{-5mm}
\end{table}
\subsubsection{CSI Feedback}
We evaluate the full-shot and zero-shot performance of WiFo-INR-Large for CSI feedback. 
Since the number and dimension of the compressed tokens are fixed at 4 and 16, respectively, the compression ratio varies across datasets with different CSI sizes. 
WiFo-CF is pretrained at a fixed compression ratio of 1/32, whereas CsiNet and TransNet are trained and tested on each dataset using the same compression ratio as WiFo-INR. 
All models adopt the same 5-bit quantization for the compressed features.
Table \ref{result-csi-feedback} reports the CSI reconstruction NMSE of all models. 
Across all 16 pretraining datasets, WiFo-INR substantially outperforms the competing methods, reducing the average NMSE by 3.28 dB and 5.63 dB compared with TransNet and CsiNet, respectively. 
On the datasets with a compression ratio of 1/32, it also achieves an average NMSE reduction of 2.46 dB over WiFo-CF. 
More importantly, WiFo-INR exhibits strong zero-shot performance on the four unseen datasets, further reducing the average NMSE by 2.36 dB compared with the full-shot performance of the strongest task-specific baseline, TransNet. 
These results demonstrate the powerful and generalizable CSI compression capability of WiFo-INR.

Moreover, Table \ref{cost-cf} compares the UE- and BS-side inference costs of different schemes for single-sample inference on the PC13 dataset at a compression ratio of 1/32. 
WiFo-INR requires more parameters and FLOPs than WiFo-CF because the latter directly processes 2D CSI without temporal padding and is dedicated solely to CSI feedback. 
Nevertheless, its inference-efficient architecture reduces the UE- and BS-side inference times by 14.43\% and 87.67\%, respectively. 
Overall, WiFo-INR achieves superior CSI feedback performance and lower inference latency than feature-based wireless foundation models tailored to CSI feedback.
\begin{table}[!t]
\footnotesize
\caption{Performance and inference cost comparison among different schemes for wireless localization. The best performance is highlighted in \textbf{bold}.}
\vspace{-2mm}
\label{WL}
\centering
\setlength{\tabcolsep}{3pt}
\begin{tabular}{cccc}
\toprule[1pt]
 & WiFo-INR & WiFo-2 & WiT \\
\midrule
RMSE (m)             & \textbf{3.93}  & 4.05             & 4.17  \\
Feature dimension    & $4\times 16$     & $64\times 384$ & /      \\
Trainable / total params (M) & 0.03 / 12.23  & 0.39 / 13.28             & 18.93 / 18.93 \\
FLOPs (G) & 3.69 & 2.74 & 2.48 \\
\bottomrule[1pt]
\end{tabular}
\vspace{-3mm}
\end{table}
\begin{table}[!t]
\footnotesize
\caption{Performance and inference cost comparison among different schemes for beam prediction. The best result is highlighted in \textbf{bold}.}
\vspace{-2mm}
\label{BP}
\centering
\setlength{\tabcolsep}{3.5pt}
\begin{tabular}{c|cccc}
\toprule[1pt]
 & WiFo-INR
 & TransNet
 & \makecell{TransNet\\$+$BP-DNN}
 & \makecell{TransNet\\$+$WiFo-2} \\
\midrule
Acc.@1
& \textbf{0.861}
& 0.677
& 0.654
& 0.857 \\

Feature dimension
& 64
& 64
& /
& 24576 \\

\makecell[c]{Trainable params (M)\\Total params (M)}
& \makecell{0.066\\12.259}
& \makecell{0.0661\\1.488}
& \makecell{1.198\\2.621}
& \makecell{1.572\\15.883} \\

BS-side FLOPs (G)
& \multicolumn{2}{c}{$0.131\times10^{-3}$}
& 0.038
& 1.404 \\
\bottomrule[1pt]
\end{tabular}
\vspace{-3mm}
\end{table}
\begin{table}[!t]
\footnotesize
\caption{Performance (F1 score) and inference cost comparison between
WiFo-INR and other baselines on scenario classification.
The CSI size is denoted as (time, subcarrier, antenna).
The best and second-best results are highlighted in
\textbf{bold} and \underline{underlined}, respectively.}
\vspace{-2mm}
\label{SC}
\centering
\begin{threeparttable}
\setlength{\tabcolsep}{4.5pt}
\begin{tabular}{c|c|ccc}
\toprule[1pt]
Dataset & CSI size & WiFo-INR & WiFo-2 & ST-CNN \\
\midrule
S1 & (8, 32, 8)
   & \underline{0.788}
   & \textbf{0.807}
   & 0.724 \\

S2 & (8, 16, 8)
   & \textbf{0.764}
   & \underline{0.718}
   & 0.710 \\

S3 & (12, 16, 8)
   & \underline{0.712}
   & \textbf{0.788}
   & 0.691 \\

S4 & (12, 32, 8)
   & \underline{0.780}
   & \textbf{0.817}
   & 0.722 \\

S5 & (4, 32, 8)
   & \textbf{0.773}
   & \underline{0.741}
   & 0.679 \\

S6 & (4, 16, 8)
   & \textbf{0.750}
   & \underline{0.652}
   & 0.631 \\
\hline

\rowcolor{gray!15}
\textbf{Avg.}
   & /
   & \textbf{0.761}
   & \underline{0.753}
   & 0.692 \\

\midrule
\makecell{S7 (full-shot)\\S7 (zero-shot)}
   & (12, 8, 8)
   & \makecell{/\\\textbf{0.662}}
   & \makecell{\textbf{0.670}\\
     \underline{0.588}\tnote{*}}
   & \makecell{\underline{0.645}\\/} \\

\midrule
\makecell{Trainable params (M)\\Total params (M)}
   & /
   & \makecell{0.016\\12.209}
   & \makecell{0.147\\13.034}
   & \makecell{4.996\\4.996} \\
\hline

\multicolumn{2}{c|}{Avg. FLOPs (G)}
   & 1.257
   & 0.477
   & 0.553 \\
\bottomrule[1pt]
\end{tabular}

\begin{tablenotes}[flushleft]
\scriptsize
\item[*] For S7, the CSI is padded along the subcarrier dimension to $(12, 16, 8)$, and zero-shot generalization is performed using WiFo-2 trained on S3.
\end{tablenotes}
\end{threeparttable}
\vspace{-3mm}
\end{table}
\begin{table}[!t]
\footnotesize
\caption{Ablation results of the first-stage pretraining on
frequency-domain channel prediction. Reconstruction NMSE results in
dB are reported. The best performances are highlighted in
\textbf{bold}.}
\vspace{-2mm}
\label{abalation-cp}
\centering
\setlength{\tabcolsep}{3.5pt}
\begin{tabular}{c|cccc}
\toprule[1pt]
 & WiFo-INR-Small
 & \makecell{w/o Fourier\\bases}
 & \makecell{w/o raw\\coordinates}
 & \makecell{SIREN\\$\rightarrow$ MLP} \\
\midrule

\makecell{\textbf{Avg.}\\(PC1--PC16)}
& \textbf{-12.38}
& \makecell{-11.09\\($\uparrow 1.29$)}
& \makecell{-11.92\\($\uparrow 0.46$)}
& \makecell{-11.63\\($\uparrow 0.75$)} \\

\makecell{\textbf{Avg.}\\(GC1--GC4)}
& \textbf{-13.88}
& \makecell{-13.00\\($\uparrow 0.88$)}
& \makecell{-13.42\\($\uparrow 0.46$)}
& \makecell{-12.98\\($\uparrow 0.90$)} \\

\bottomrule[1pt]
\end{tabular}
\vspace{-3mm}
\end{table}
\begin{table*}[!t]
\footnotesize
\caption{Ablation results of the second-stage pretraining on CSI feedback. Reconstruction NMSE results in dB are reported. The best performances are highlighted in \textbf{bold}.}
\label{ablation-cf}
\centering
\begin{tabular}{c|ccccc}
\toprule[1pt]
 & WiFo-INR-Large
 & Full training
 & \makecell{Training from scratch}
 & \makecell{w/o progressive compression}
 & w/o pre/post MLPs\\
\midrule
Avg. (PC1-PC16)
& \textbf{-13.26}
& -12.84 ($\uparrow 0.42$)
& -9.43 ($\uparrow 3.83$)
& -12.56 ($\uparrow 0.70$)
& -11.74 ($\uparrow 1.52$) \\

Avg. (GC1-GC4)
& -15.13
& \textbf{-15.26} ($\downarrow 0.13$)
& -10.19 ($\uparrow 4.94$)
& -14.30 ($\uparrow 0.83$)
& -13.44 ($\uparrow 1.69$) \\

Trainable params (M)
& 1.32
& 12.19
& 12.19
& 1.26
& 0.90 \\
\bottomrule[1pt]
\end{tabular}
\vspace{-3mm}
\end{table*}
\subsubsection{CSI-Related Downstream Tasks}

For wireless localization, Table \ref{WL} reports the localization accuracy and computational costs of WiFo-INR-Large and baselines, with WiFo-INR achieving SOTA accuracy. 
Although WiFo-INR incurs slightly higher FLOPs, it reduces the number of trainable parameters by 92.31\% compared with WiFo-2. 
This is because WiFo-INR fine-tunes a lightweight regression head based on compact modulation tokens, whereas the feature dimension of WiFo-2 is much higher.
These results demonstrate that the compact representations of WiFo-INR can not only be effectively transferred to downstream regression tasks but also significantly reduce fine-tuning overhead.

For beam prediction, we consider the following baselines. 
The TransNet-based baseline is fine-tuned using the compressed features received at the BS, with WiFo-INR replaced by TransNet pretrained on the target dataset. 
The other two baselines operate on the CSI reconstructed by TransNet: BP-DNN processes the reconstructed CSI directly, while WiFo-2 first extracts CSI features and then applies an output head. 
Table \ref{BP} reports the prediction accuracy and inference costs of different schemes.
Compared with these reconstruction-based schemes, directly processing compressed features substantially reduces the BS-side fine-tuning overhead, feature dimension, and number of trainable parameters. 
Moreover, using compressed features of the same dimensionality, WiFo-INR achieves a 27.2\% relative improvement in classification accuracy over TransNet, further demonstrating its strong representation capability.

We further evaluate scenario classification to assess the generalization of WiFo-INR-Large across CSI sizes. 
Since the dimension of the modulation token is independent of the CSI size, WiFo-INR can be jointly trained on S1-S6 with a shared output head and directly generalized to the unseen S7 size. 
In contrast, WiFo-2 requires a separate head for different CSI sizes.
It must either be fine-tuned on S7 or perform zero-shot inference by padding S7 to the S3 size, while ST-CNN must be retrained for each dataset. 
As shown in Table \ref{SC}, WiFo-INR achieves the highest average F1 score on S1-S6, outperforming the second-best method by 1.06\%.
On S7, its zero-shot F1 score exceeds that of WiFo-2 by 12.59\% and even surpasses the full-shot ST-CNN. 
It also reduces the number of trainable parameters by 89.12\% compared with WiFo-2, although its average FLOPs are higher due to the additional learnable tokens. 
Overall, its compact representation enables effective generalization across CSI sizes.
\subsection{Ablation Experiments}
To evaluate the effectiveness of the key modules and design choices, we conduct two groups of ablation experiments. 
For the first-stage pretraining, we remove the Fourier bases or raw coordinates from the Fourier features in \eqref{fourier}, and replace all SIREN layers with MLP layers. 
Table \ref{abalation-cp} reports the frequency-domain channel prediction results based on WiFo-INR-Small. 
Since these variants have negligible impact on the parameter count, the corresponding values are omitted. 
All ablations degrade performance on both the pretraining and generalization datasets, confirming the effectiveness of the designed Fourier features and SIREN architecture in enhancing the modeling capability.

For the second-stage pretraining, we compare the default LoRA-based scheme with full-parameter training and training from random initialization. 
We also replace the two-stage progressive compression with single-stage compression and remove the pre- and post-compression MLPs.
As these variants substantially affect the training cost, Table \ref{ablation-cf} reports both the CSI feedback performance and the number of trainable parameters of WiFo-INR-Large. 
All ablations degrade the average NMSE to varying degrees. 
The default scheme achieves performance comparable to full-parameter training while using only 10.83\% of the trainable parameters, demonstrating the effectiveness of LoRA-based adaptation. 
Training from scratch increases the NMSE by more than 3.8 dB, highlighting the importance of first-stage pretraining. 
Single-stage compression and MLP removal further reduce the trainable parameters but lead to substantial performance degradation. 
Overall, the default design offers the best tradeoff between task performance and training cost.
\subsection{Scaling Analysis}
\begin{figure}[!t]
    \centering
    \subfloat[Scaling analysis on the model scale.]{
        \includegraphics[width=0.9\linewidth]{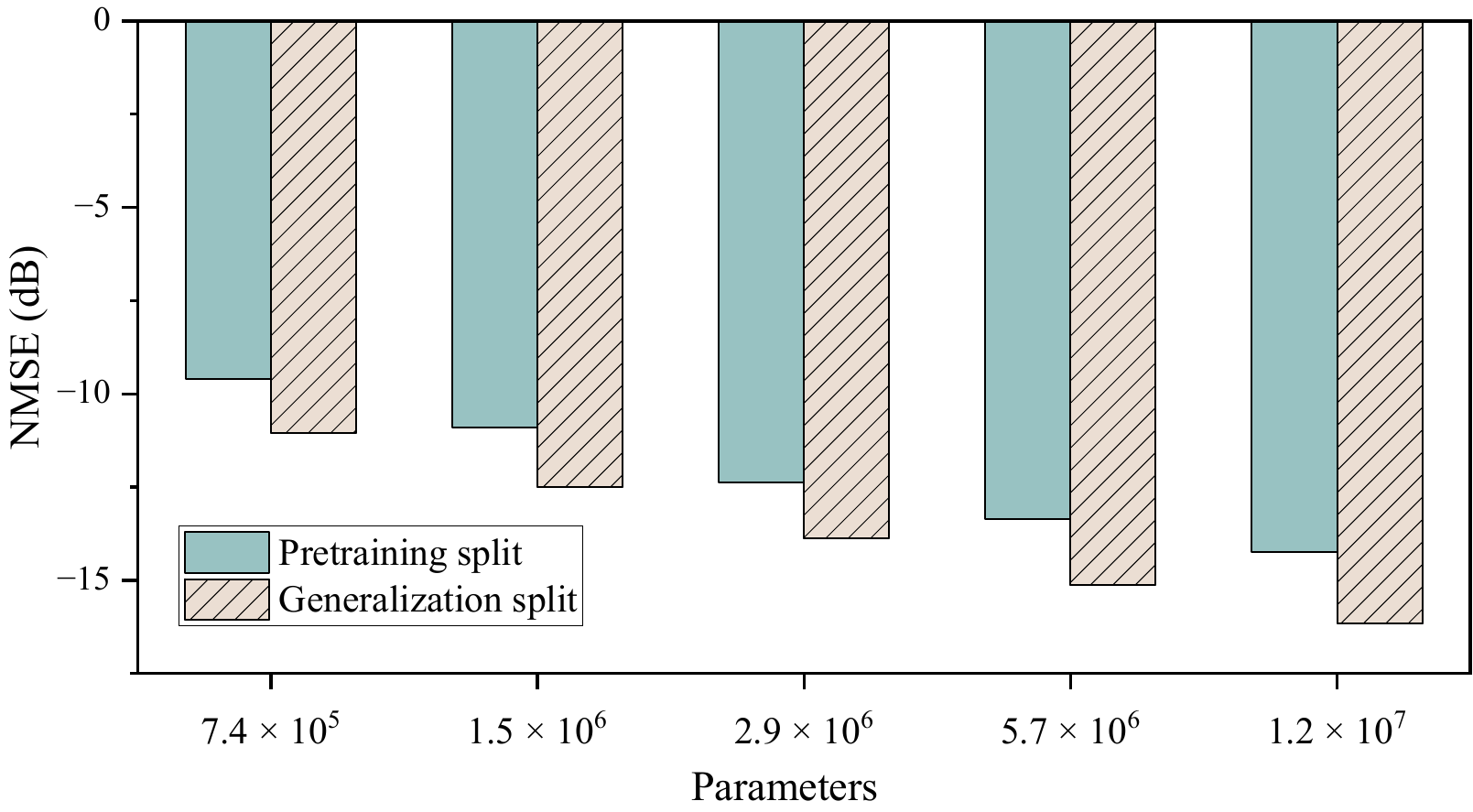}
        \label{model-scale}
    }
    \vspace{2mm}
    \subfloat[Scaling analysis on the data scale.]{
        \includegraphics[width=0.9\linewidth]{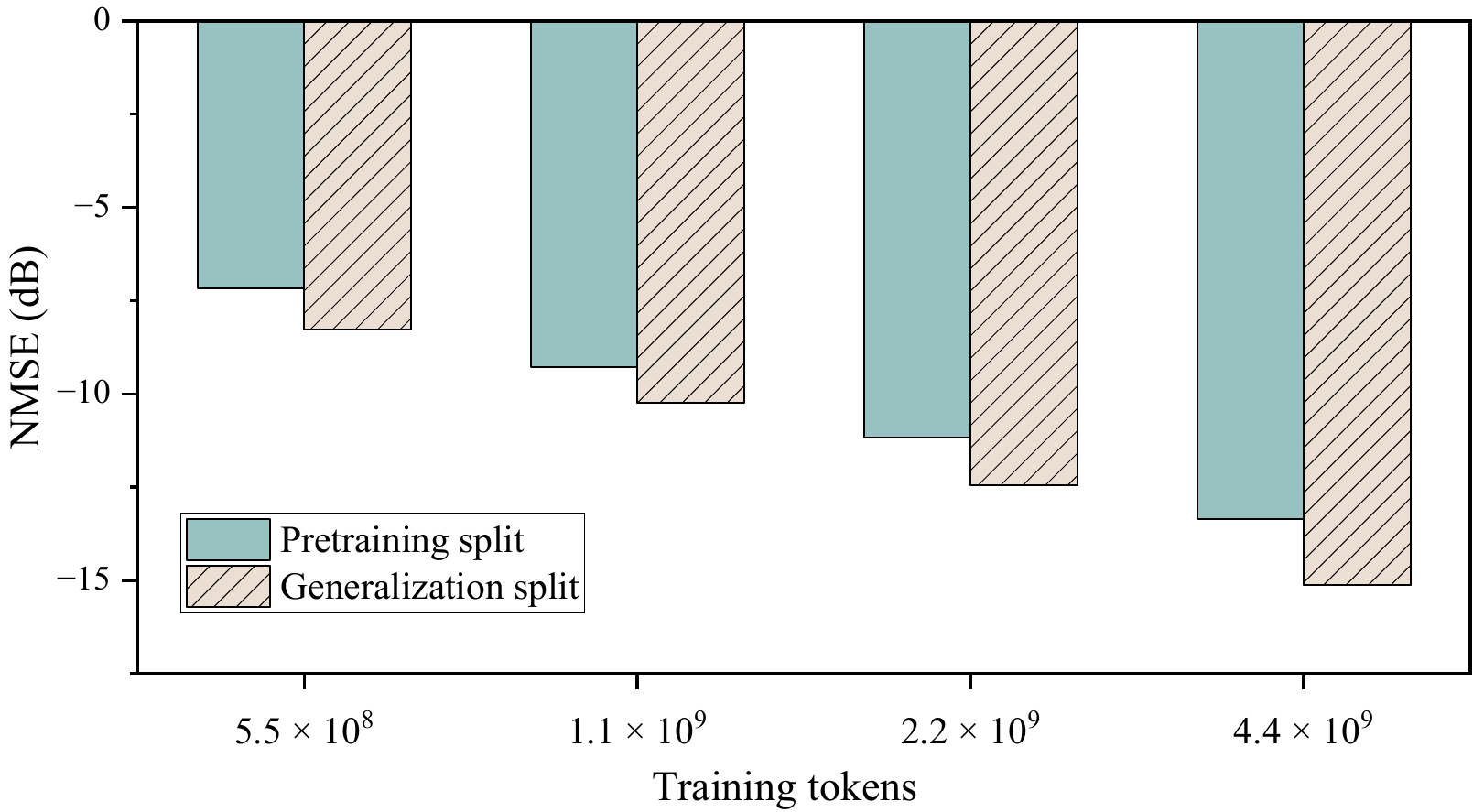}
        \label{data-scale}
    }
    \caption{Scaling analysis of frequency-domain channel prediction performance with respect to model and data scale.}
    \vspace{-5mm}
\end{figure}
We further investigate the empirical scaling behavior of WiFo-INR with respect to model and data scales, using the average test NMSE for frequency-domain channel prediction on both the pretraining and generalization-validation datasets as the evaluation metric. For model scaling, we use all pretraining datasets and evaluate five WiFo-INR variants, ranging from Tiny to Large, as shown in Fig. \ref{model-scale}. For data scaling, we fix the model architecture to WiFo-INR-Base and vary the number of training tokens, as shown in Fig. \ref{data-scale}.
For both the pretraining and generalization splits, the NMSE consistently decreases as the model size and the amount of pretraining data increase, indicating favorable empirical scalability of WiFo-INR over the evaluated range. These results provide practical guidance for selecting an appropriate model size and pretraining data scale based on the target performance, available pretraining compute budget, and the deployment platform's hardware constraints.

\section{Conclusion}
In this paper, we proposed WiFo-INR, an INR-based wireless foundation model that uses coordinate-conditioned neural functions for CSI modeling. 
Beyond existing methods that rely on per-instance optimization, we designed an architecture that generalizes across CSI instances.
It comprised a Transformer-based encoder for modulation-token generation and a SIREN-based decoder serving as the INR. 
We further adopted a two-stage pretraining strategy to enhance channel reconstruction and CSI feedback. 
Experiments demonstrated that the compact and efficient implicit representation offered clear advantages over existing wireless foundation models based on explicit discrete modeling. 
For channel reconstruction, WiFo-INR outperformed WiFo-2 while substantially reducing inference cost. 
For CSI feedback, it achieved better performance and lower inference time than WiFo-CF. 
For CSI-related downstream tasks, its compact representations maintained strong task performance while substantially reducing fine-tuning overhead and enabling generalization across CSI sizes. 
Ablation studies validated the effectiveness of the proposed modules, while the scaling analysis demonstrated the scalability of the architecture.
\bibliographystyle{IEEEtran}
\small
\bibliography{IEEEabrv, ref}

\end{document}